\documentclass[usenatbib,usegraphicx,letter]{mn2e}

\newcommand{\LG} {P/LG}

\title[Comet P/2010~TO20 LINEAR-Grauer as a Mini-29P/SW1]{Comet
P/2010~TO20 LINEAR-Grauer as a Mini-29P/SW1}
\author[Pedro Lacerda]{Pedro Lacerda$^{1}$\thanks{E-mail:
  p.lacerda@qub.ac.uk}\thanks{Some of the presented
  data were obtained at the European Southern Observatory (ESO)
facilities at La Silla under programme 088.C-0634A.}\\
$^{1}$Astrophysics Research Centre, School of Mathematics and Physics,
Queen's University Belfast, Belfast BT7 1NN}
\begin{document}

\date{Accepted for publication on MNRAS.\\ Accepted 2012 October 4.  Received 2012 October 4; in original form 2012 July 24}

\pagerange{\pageref{firstpage}--\pageref{lastpage}} \pubyear{2012}

\maketitle

\label{firstpage}

\begin{abstract}

Discovered in October 2010 by the LINEAR survey, P/2010 TO20
LINEAR-Grauer (\LG) was initially classified as an inert Jupiter
Trojan.  Subsequent observations obtained in October 2011 revealed
\LG\ to be a Jupiter-family comet. \LG\ has one of the largest
perihelia ($q=5.1$ au) and lowest eccentricities ($e=0.09$) of the
known JFCs. We report on observations of \LG\ taken on 29 October 2011
and numerical simulations of its orbital evolution. Analysis of our
data reveals that \LG\ has a small nucleus ($<3$ km in radius) with
broadband colours ($B-R=0.99\pm0.06$ mag, $V-R=0.47\pm0.06$ mag)
typical of JFCs. We find a model dependent mass-loss rate close to 100
kg s$^{-1}$, most likely powered by water-ice sublimation. Our
numerical simulation indicate that the orbit of \LG\ is unstable on
very short (10 to 100 yr) timescales and suggest this object has
recently evolved into its current location from a more distant,
Centaur-type orbit. The orbit, dynamics and activity of \LG\ share
similarities with the well known case of comet
29P/Schwassmann-Wachmann 1. 

\end{abstract}


\section{Introduction}

Jupiter family comets (JFCs) originate in the transneptunian region
of the solar system known as the Kuiper belt. Kuiper belt objects
(KBOs) preserve key information about the epoch of planetesimal
formation. The low ambient temperature ($\sim40$ K) allows KBOs to
retain most if not all of the ices present in their formation
environment. This makes them valuable time capsules with which to
study an epoch long gone. JFCs represent the small end of the steep
size distribution of KBOs. They are survivors of a dynamically
intermediate population, the Centaurs, that were neither ejected from
the solar system nor collided with one of the giant planets (Jupiter
to Neptune) on their journey from the Kuiper belt into the inner solar
system.

The surfaces of JFCs are heavily processed when compared to KBOs,
mainly by sublimation of surface ice. Indeed, the optical broadband
colours of the two populations differ substantially: JFCs have
typically solar (neutral) colours ($B-R\sim1.3$, $g-r\sim0.6$), with a
spread of about 0.2 mag \citep{2009Icar..201..674Lamy,
2012Icar..218..571Solontoi} while KBOs span a broad range of optical
colours, from neutral ($B-R\sim1$) to very red
\citep[$B-R\sim2.5$,][]{2001AJ....122.2099J, 2002AJ....123.1039Jew}.
The neutral surfaces are usually attributed to a fresh ice coating
while very red surfaces are thought to be the end product of
irradiation of initially neutral material
\citep{1987JGR....9214933Thompson, 1998Icar..134..253Moroz}.  The
intermediate Centaurs present a peculiar, also intermediate
distribution of colours which includes neutral objects (blue group)
and very red objects (red group) but very few cases in between
\citep{2003A&A...410L..29Pei}. This Centaur colour bimodality may be
related to their dynamical history \citep{2012A&A...539A.144Melita} or
simply to the fact that they are small
\citep{2012arXiv1206.3153Peixinho}.

Some objects straddle the line between Centaurs and JFCs. They are
active, like JFCs, but have Centaur-like orbits. A famous example is
29P/Schwassmann-Wachmann 1 (hereafter 29P). The orbit of 29P is nearly
circular ($e=0.04$) with perihelion and semimajor axis beyond Jupiter
($q=5.7$ au and $a=6.0$ au), and obeys the dynamical definition of a
Centaur.  29P is constantly active and displays sporadic photometric
outbursts \citep{1958PASP...70..272Roemer, 1990ApJ...351..277Jewitt,
2008A&A...485..599TrigoRodriguez}.  The activity of 29P is mainly
driven by carbon monoxide sublimation \citep{1994Natur.371..229Senay,
1995Icar..115..213Crovisier}. Comet 29P has a radius $r=23\pm3$ km
(Yan Fern\'andez, private comm.).

\begin{figure}
  \includegraphics[width=83mm]{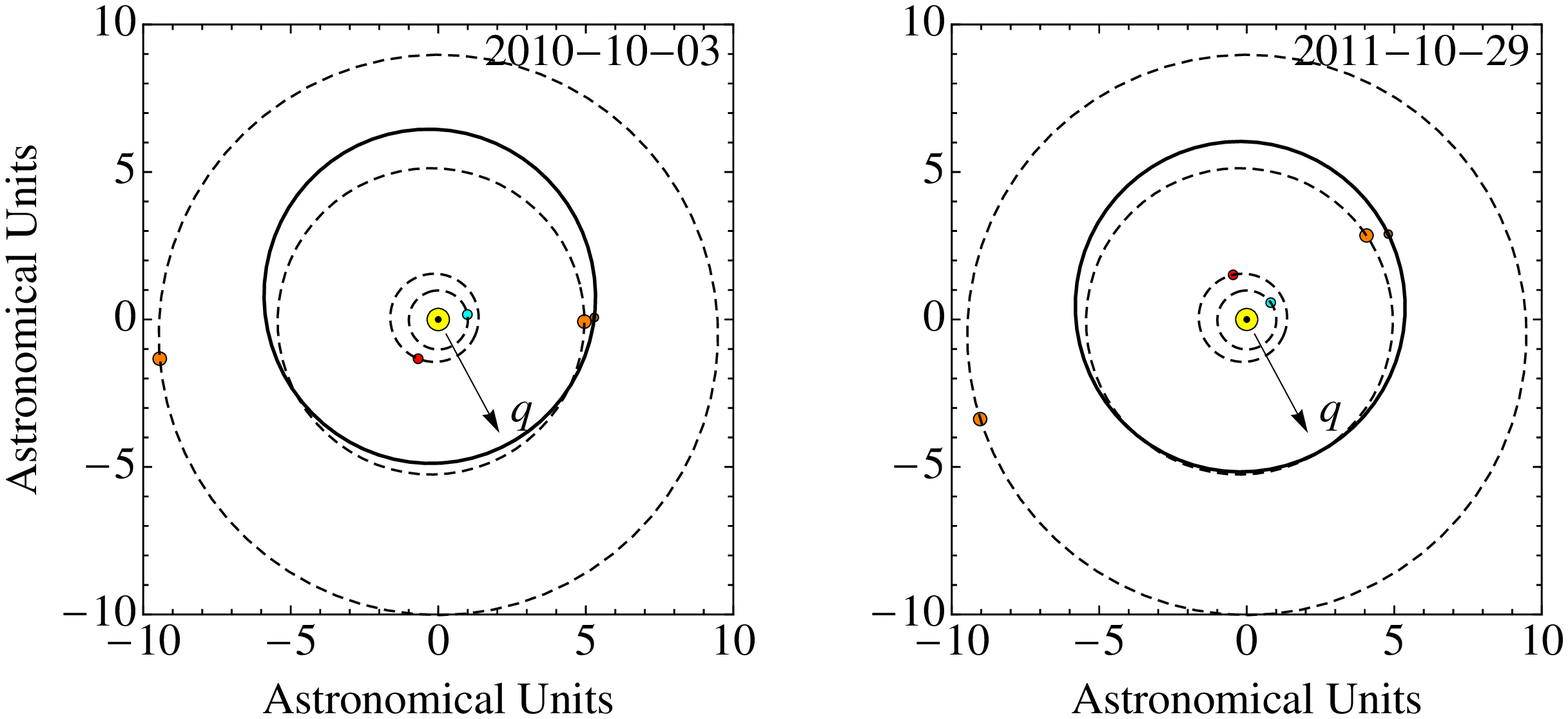}
  \centering

  \caption{Osculating orbit of \LG\ on 3 October 2010 (near the time it
    was discovered by LINEAR) and on 29 October 2011, when the
    observations reported here were taken. Due to the proximity to
    Jupiter the osculating orbit of \LG\ changed significantly in a
    period of only one year. An arrow marks the direction to the last
    perihelion, the orbits of Earth, Mars, Jupiter and Saturn are
    plotted as dashed lines and the axes are labelled in astronomical
  units.}

  \label{Fig.Orbit}
\end{figure}

In this paper we present a study of comet P/2010~TO20 LINEAR-Grauer
(hereafter \LG) which is in many respects similar to 29P. Comet \LG\
has a low eccentricity ($e=0.09$), high perihelion ($q=5.1$ au) orbit,
very close to meeting the Centaur definition (see Table
\ref{Table.Elements}, Fig.\ \ref{Fig.Orbit}). When \LG\ was discovered
by the Lincoln Near Earth Asteroid Research (LINEAR) survey, on 1
October 2010, it was misclassified as a Trojan and its weak activity
went unnoticed. We present optical broadband measurements of \LG\ and
numerical simulations of its orbital evolution. The former show that
\LG\ is weakly active, at a level comparable to the active Centaurs
\citep{2009AJ....137.4296Jewitt} while the latter indicate that \LG\
has in all likelihood recently evolved into its current location from
a Centaur orbit. The implication is that \LG\ may offer a glimpse of a
relatively small and fresh Centaur object. 

\begin{figure*}
  \includegraphics[width=140mm]{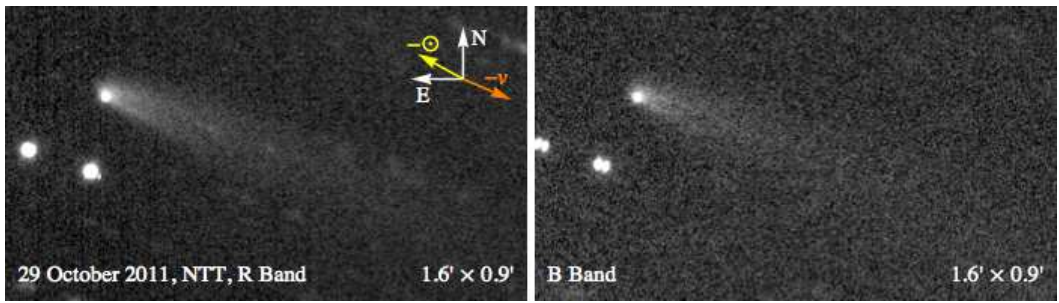}
  \centering

  \caption{Summed stacks of 9 R-band (left) and 6 B-band
      (right) images of comet \LG. The projected tail is more than
      1\arcmin\ long and points in the direction opposite the orbital
      motion ($-v$).  Because of the very small solar phase angle
      (Table \ref{Table.Geometry}), the antisolar direction lies
      nearly perpendicularly to the plane of the sky. For this reason,
      the near-nucleus tail has only a small component in the
      projected antisolar direction ($-\sun$; see Fig.\
      \ref{Fig.FilteredInsets}) which rapidly bends towards the
    direction opposite the orbital motion.  }

    \label{Fig.Stacks}
\end{figure*}

\begin{figure*}
  \includegraphics[width=140mm]{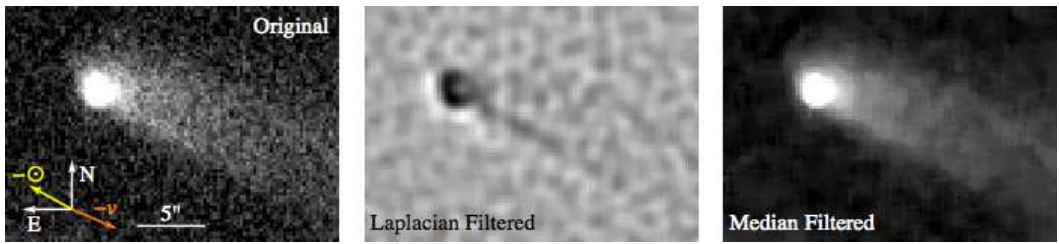}
  \centering

  \caption{Enlarged versions of the near-nucleus
      tail structure from the R-band stack shown in Fig.\
      \ref{Fig.Stacks}.  The original image (left) is shown together
      with versions processed using Laplacian (centre) and median
      (right) filters. A trail of larger particles leaving the nucleus
      and trailing the comet's orbital motion is visible in the
      Laplacian-filtered image. The median-filtered image highlights
      the near-nucleus part of the tail which points in the projected
    antisolar direction. }

    \label{Fig.FilteredInsets}
\end{figure*}

\begin{table}
  \caption{Orbital properties of LINEAR-Grauer on 2011 Oct 29. \label{Table.Elements}}
  \begin{tabular}{@{}lr@{}}
    \hline
    Property & Value \\
    \hline
    Semimajor axis, $a$                    & 5.610 au \\
    Eccentricity, $e$                      & 0.087  \\
    Inclination, $i$                       & 2$\fdg$628  \\
    Argument of perihelion, $\omega$       & 252$\fdg$9  \\
    Longitude of ascending node, $\Omega$  & 44$\fdg$03  \\
    Mean anomaly, $M$                      & 84$\fdg$22  \\
    True anomaly, $\nu$                    & 94$\fdg$19 \\
    Last perihelion passage                & 2008 Sep 5 \\
    Perihelion distance, $q$               & 5.122 au \\
    Aphelion distance, $Q$                 & 6.097 au \\
    \hline
  \end{tabular}
\end{table}

\begin{table}
  \caption{Observing Geometry. \label{Table.Geometry}}
  \begin{tabular}{@{}lrr@{}}
    \hline
    Property & 2010 October 03 & 2011 October 29 \\
    \hline
    Solar phase angle ($\alpha$)            & 2.07\degr & 0.89\degr\\
    Heliocentric distance ($R$)             & 5.295 au  & 5.603 au \\
    Geocentric distance ($\Delta$)          & 4.309 au  & 4.613 au \\
    1\arcsec\ at distance $\Delta$          & 3135 km   & 3355 km  \\ 
    \hline
  \end{tabular}
\end{table}

\begin{table*}
  \centering
  \begin{minipage}{160mm}
    \caption{Photometry Regions \label{Table.PhotometryRegions}}
    \begin{tabular}{@{}ccccccc@{}}
      \hline
      Region & Region & Aperture radius & Aperture radius  & Projected
      radius & Magnitudes & Dominant Source \\
      Label  & Shape  & [pixels] & [\arcsec] & [km] & in Region & \\
      \hline
      $\mathcal{R}_{4}$     & Circle  & $r=4$        & $\phi=0.96$          & $d=3,220$            & $m_4$      & Nucleus (+ Coma) \\
      $\mathcal{R}_{4,8}$   & Annulus & $r=4$ to 8   & $\phi=0.96$ to 1.92  & $d=3,220$ to 6,440   & $m_{4,8}$  & Nucleus + Coma   \\
      $\mathcal{R}_{8,13}$  & Annulus & $r=8$ to 13  & $\phi=1.92$ to 3.12  & $d=6,440$ to 10,465  & $m_{8,13}$  & Coma            \\
      $\mathcal{R}_{13,20}$ & Annulus & $r=13$ to 20 & $\phi=3.12$ to 4.80  & $d=10,465$ to 16,100 & $m_{13,20}$ & Coma      \\
      \hline
    \end{tabular}
  \end{minipage}
\end{table*}

\section[]{Observations} \label{Sec.Observations}

\LG\ was observed on 29 October 2011 at the ESO New Technology
Telescope (NTT) located at the La Silla Observatory, Chile.  The night
was photometric and the seeing varied between 0.8 and 1.0\arcsec. At
the NTT we used the EFOSC2 instrument
\citep{1984Msngr..38....9Buzzoni, 2008Msngr.132...18Snodgrass} which
is installed at the f/11 Nasmyth focus and is equipped with a LORAL
$2048\times2048$ CCD. We used the $2\times2$ binning mode to bring the
effective pixel scale to 0.24\arcsec/pixel. Our observations were
taken through Bessel $B,V,R$ filters (ESO \#639, \#641, \#642,
respectively).

The images of \LG\ were collected in a relatively regular fashion, in
sets of three consecutive exposures per filter, with exposure times of
240 s for $B$ and 120 s for $V$ and $R$. In total, we collected 12
images in the $B$ band, 12 in $V$, and 18 in $R$. Throughout the
observations of \LG, the telescope was set to track the non-sidereal
motion of \LG\ at an approximate rate of $-18$\arcsec/hour in right
ascension and $-6$\arcsec/hour in declination.

Bias calibration frames and dithered twilight flats through all three
filters were collected on the same night as the science data. The
reduction of the science images, consisting of standard bias
subtraction and flat fielding, was done using the IRAF {\tt ccdproc}
routines.  The R band images suffered from fringing which was removed
using an IRAF package optimised for EFOSC2 that was kindly supplied by
Colin Snodgrass. The \LG\ data were absolutely calibrated using
observations of \citet{1992AJ....104..340Lan} stars taken throughout
the night. 

\LG\ was observed as part of the Pan-STARRS PS1 all-sky survey very
near the time of its discovery by LINEAR. Located on Haleakala, Maui,
the 1.8-m PS1 telescope is equipped with a 1.4 gigapixel camera
covering $3.2\degr\times3.2\degr$ on the sky. The survey repeatedly
covers the $3\pi$ steradians of sky visible from Haleakala.  PS1 uses
a photometric system that approaches the SDSS filter system with the
addition of a wide ($w$) band filter which roughly corresponds to the
combined pass band of the $gri$ filters
\citep{2012ApJ...750...99Tonry}. The PS1 gigapixel camera has a pixel
scale 0.25\arcsec/pixel.  \LG\ was imaged in four consecutive 45 s
exposures taken through the $w$ filter on 3 October 2010. All images
were processed automatically by the Pan-STARRS Image Processing
Pipeline.

\section{Comet Morphology} \label{Sec.Coma}

Figure \ref{Fig.Stacks} shows R- and B-band summed stacks of \LG\ and
Fig.\ \ref{Fig.FilteredInsets} shows enlarged sections of the region
surrounding the nucleus of \LG, to highlight details. A description of
how the stacks were assembled can be found in \S\ref{Sec.NucleusSize}
and \S\ref{Sec.NucleusColour}; here we focus solely on inspecting the
overall appearance of the comet.  The projected comet tail is clearly
visible and extends for more than 1\arcmin\ in the direction opposite
the orbital motion.  Due to the very small solar phase angle, the
antisolar direction points almost radially away from the observer.  As
a result, the projected, antisolar component of the cometary tail is
tiny, and more easily viewed in the enlarged, median-filtered version
shown in Fig.\ \ref{Fig.FilteredInsets}. The bulk of the tail
particles trails the comet as a result of Keplerian shear. The
Laplacian-filtered image (Fig.\ \ref{Fig.FilteredInsets}) displays a
darker ridge at the centre of the tail which originates in the nucleus
and is aligned with the orbital motion. The ridge is probably due to
large particles leaving the nucleus.  Larger particles attain lower
ejection velocities as they couple more weakly to the sublimating gas
and hence tend to remain concentrated in the plane of the orbit where
solar radiation acts to push them radially outwards and Keplerian
shear forces them to trail the nucleus.

\begin{figure}
  \includegraphics[width=83mm]{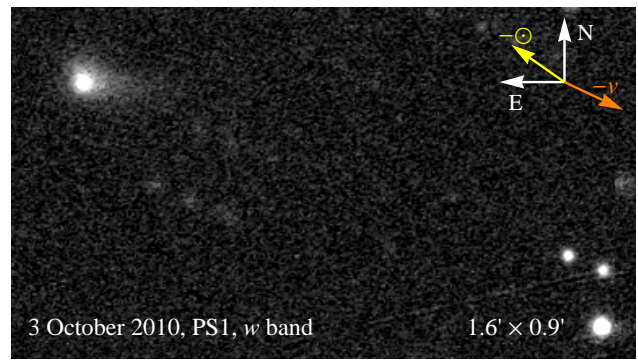}
  \centering

  \caption{Stack of four consecutive PanSTARRS $w$-band
      images of \LG\ taken on 3 October 2010, very near the time of its
      discovery by the LINEAR survey.  The comet is clearly active in
    these images.} 

    \label{Fig.PSOne}
\end{figure}

Figure \ref{Fig.PSOne} shows a stack of four consecutive $w$-band
images of comet \LG\ taken by the PS1 survey. The PS1 images are taken
only 2 days after discovery by LINEAR on 1 October 2010. The comet
appears active, with a tail that extends about 12\arcsec\ on the sky.
Here, as in the case of our more recent observations, the solar phase
angle is small, just over 2\degr, probably causing the tail to extend
mainly into the plane of the sky along the line of sight. As above,
the near-nucleus portion of the sky-projected tail seems to be
antisolar, but further from the nucleus the tail seems to trail the
nucleus along the orbital direction. The three sets of images show that
\LG\ was active at the time of discovery (October 2010) and remains
active a year later (October 2011).

\section{Photometry} \label{Sec.Photometry}

We performed aperture photometry of comet \LG\ and nearby field stars
using the IRAF {\tt apphot} task with four main goals: 1) to estimate
the size of the comet nucleus, 2) to estimate the mass-loss rate from
the comet, 3) to measure the colour of the nucleus and coma dust, and
4) to search for brightness variations in the nucleus and coma over
time. With those goals in mind, we extracted photometric data from the
science images using four synthetic apertures defining four regions of
interest (see Table \ref{Table.PhotometryRegions}). The central
region, $\mathcal{R}_4$, 4 pixels ($0.96\arcsec$) in radius, was
selected to represent the flux from the nucleus, but includes an
unknown fraction of light due to the coma. The intermediate annulus,
$\mathcal{R}_{4,8}$, between 4 and 8 pixels ($0.96\arcsec$ and
$1.92\arcsec$) from the nucleus includes contributions from both the
nucleus point spread function (PSF) and the coma.  The annular region
$\mathcal{R}_{8,13}$, located between 8 and 13 pixels ($1.92\arcsec$
and $3.12\arcsec$) from the nucleus, is dominated by the coma.
Finally, the outer annulus $\mathcal{R}_{13,20}$, located between 13
and 20 pixels from the nucleus ($3.12\arcsec$ and $4.8\arcsec$),
includes coma flux and significant noise contamination.  Details of
how these particular apertures were chosen can be found in
\S\ref{Sec.NucleusSize}, \S\ref{Sec.MassLoss} and
\S\ref{Sec.NucleusColour}.  Magnitudes measured in these regions are
labeled $m_4$, $m_{4,8}$, $m_{8,13}$, and $m_{13,20}$.  Magnitudes
within the annuli, labeled $m_\mathrm{in,out}$, were calculated from
the magnitudes within the inner and outer apertures, $m_\mathrm{in}$
and $m_\mathrm{out}$, through \[ m_\mathrm{in,out}=-2.5 \log_{10}
\left(10^{-0.4 m_\mathrm{out}}-10^{-0.4 m_\mathrm{in}}\right).  \]
\noindent We list the $B$, $V$ and $R$ apparent magnitudes in each
region in Table \ref{Table.Photometry}.

\begin{table}
  \caption{Photometry \label{Table.Photometry}}
  \begin{tabular}{@{}lcccc@{}}
    \hline
    Region   & $B$ band & $V$ band & $R$ band \\
    measured & [mag]    & [mag]    & [mag]    \\
    \hline
    $m_4$       & $21.94\pm0.05$ & $21.42\pm0.05$ & $20.95\pm0.03$ \\
    $m_{4,8}$   & $22.31\pm0.08$ & $21.66\pm0.08$ & $21.21\pm0.05$ \\
    $m_{8,13}$  & $22.56\pm0.16$ & $21.80\pm0.13$ & $21.33\pm0.08$ \\
    $m_{13,20}$ & $22.72\pm0.30$ & $22.04\pm0.27$ & $21.47\pm0.15$ \\
    \hline
  \end{tabular}
\end{table}

\subsection{Nucleus Size}\label{Sec.NucleusSize}

To estimate the radius of the nucleus of comet \LG\ we began by
generating a summed stack of 9 $R$-band frames that were shifted and
aligned at the comet nucleus position. The stack, displayed in Fig.\
\ref{Fig.Stacks}, has an equivalent integration time of 1080 s and
offers an enhanced signal-to-noise ratio (SNR) detection of the comet.
We used the $R$-band images because they are sharper (better seeing)
and have higher SNR than those in $B$ and $V$. The particular frames
that were stacked have seeing between 0.8\arcsec\ and 0.9\arcsec. An
analogous stack that aligns the field stars instead of the comet was
used to investigate the best choice of aperture for the photometry. We
found that a 4-pixel radius aperture, containing $\sim85\%$ of the
flux of a point source, offers the best compromise between maximising
the SNR within the aperture and minimising the contribution from
surrounding sources.  We labeled this region $\mathcal{R}_4$ and used
the magnitude within it, $m_4$, as best representing the flux from the
nucleus.  In the case of comet \LG\ we obtain a $\mathrm{SNR}_R\sim75$
within the central region $\mathcal{R}_4$ and measure an $R$-band
magnitude $m_4=20.95\pm0.03$ mag. 

Assuming a spherical cometary nucleus with cross-section $\pi r_e^2$,
we calculated the equivalent radius $r_e$ using the relation
\citep{1916ApJ....43..173Russell}: 
\begin{equation}
  \pi r_e^2 \, p_R \, 10^{-0.4\beta\alpha} = 2.25\times10^{22} \, \pi
  \, R^2 \, \Delta^2 \, 10^{-0.4(m_4-m_{\sun})}  \label{Eq.Russell}
\end{equation}
\noindent where $p_R$ is the $R$-band geometric albedo and $\beta$ is
the linear phase coefficient of the nucleus. Table
\ref{Table.Geometry} lists the values of the solar phase angle,
$\alpha$, heliocentric distance, $R$, and geocentric distance,
$\Delta$, of \LG\ on 29 October 2011. The apparent red magnitude of the
Sun at Earth is $m_{\sun}=-27.1$ mag.  The values of $p_R$ and $\beta$
are unknown for \LG. The uncertainty introduced by $\beta$ is
negligible when compared to that in $p_R$ (uncertain by a factor 2 or
more) because \LG\ was observed at very low phase angle
($\alpha<1\degr$).  We used $\beta=0.02$ mag/\degr
\citep{1982AJ.....87.1310Millis,1987A&A...187..585Meech} and $p_R=0.1$
\citep{2004come.book..577Kolokolova} and found an equivalent radius
$r_e<3$ km. This figure is an upper limit due to the unknown
contribution of dust coma to the flux within the central aperture.

The apparent magnitude of \LG\ may be converted to a nucleus absolute
magnitude, $m(1,1,0)$ (the theoretical magnitude of the nucleus when
placed at 1 au from the Sun and the Earth, and seen at 0\degr\ phase
angle), using: \begin{equation}
  m(1,1,0)=m_4-5\log_{10}(R\Delta)-\beta\alpha.
\end{equation} Substituting the quantities $R$, $\Delta$ and $\alpha$
from Table \ref{Table.Geometry} and again using $\beta=0.02$
mag/\degr\ we obtain $m_R(1,1,0)=13.04\pm0.03$ mag. The formal error
in $m(1,1,0)$ is the same as that in the apparent red magnitude, but
the true uncertainty is higher due to the unknown value of $\beta$. We
also note that this is a lower limit to the absolute magnitude of the
\LG\ nucleus because of the unknown contribution of near-nucleus coma
to the magnitude $m_4$.

\subsection{Mass-loss rate}\label{Sec.MassLoss}

An order of magnitude estimate of the mass-loss rate from \LG\ can be
calculated by dividing the total dust mass present within a
coma-dominated annulus surrounding the nucleus by the time it takes
the dust to cross the annulus \citep{1989AJ.....97.1766Jewitt}. As in
\S\ref{Sec.NucleusSize} we used the 9-frame, $R$-band stack to measure
the amount of sunlight-reflecting dust within the annular region.  The
annulus was set between 8 and 13 pixels from the nucleus photocentre
and labeled $\mathcal{R}_{8,13}$ (Table\
\ref{Table.PhotometryRegions}).  The annulus dimensions were found by
experimentation (using the profiles of field stars in the star-centred
stack) to ensure that the flux within the annulus is dominated by the
coma and sufficient to overcome the noise due to the sky background.
In our stacked data, only 3\% of the flux of a point source is
present in the wings of the PSF beyond an aperture of 8 pixels. 

The $R$-band magnitude within the annulus $\mathcal{R}_{8,13}$ is
$m_{8,13}=20.95\pm0.03$ mag (see Table \ref{Table.Photometry}). The
magnitude $m_{8,13}$ can be converted to a total dust cross-section,
$A_d$, using Eq.\ \ref{Eq.Russell} if we use the latter instead of the
nucleus cross-section $\pi r_e^2$.  In this case, $p_R$ and $\beta$
are the $R$-band albedo and linear phase coefficient of the dust
particles, also unknown for \LG. We assumed that these values are the
same as for the nucleus ($\beta=0.02$ mag/\degr\ and $p_R=0.1$) and
found a total dust cross-section area $A_d\sim2\times 10^7$ m$^2$.
This figure is uncertain by at least a factor of 2 mainly due to the
uncertainty in the albedo.

To convert the dust cross-section, $A_d$, to a dust mass, $M_d$, we
assume that the dust consists of equal-sized spheres with an
equivalent radius, $r_d$, representative of their true size
distribution \citep[e.g.,][]{2011ApJ...728...31Li}. The mass is then
given by 
\begin{equation}
M_d = (4/3) \rho_d r_d A_d
\end{equation}
\noindent where $\rho_d$ is the bulk density of the dust particles. We
take an equivalent dust particle radius of $r_d=\left(0.1\,\mu
\mathrm{m}\times1\,\mathrm{cm}\right)^{1/2}\approx32$ $\mu$m which is
based on a power-law dust size distribution proportional to (dust
grain radius)$^{-3.5}$ with minimum and maximum grain radii of 0.1
$\mu$m and 1 cm \citep{2009AJ....137.4296Jewitt,
2011ApJ...728...31Li}. We further assume a dust bulk density
$\rho_d=1000$ kg m$^{-3}$ and find a total dust mass within the
annulus $M_d\sim8.6\times 10^5$ kg.

The annulus crossing time, $t_\mathrm{cross}$, is obtained by dividing
the annulus projected width, $w_\mathrm{cross}=4\times10^6$ m (see
Table \ref{Table.PhotometryRegions}), by the dust velocity, $v_d$. The
velocity $v_d$ is highly uncertain and depends on grain size
\citep[][]{2004come.book..471Crifo} but estimates based on macroscopic
fragment ejection from 17P/Holmes \citep{2010AJ....139.2230Stevenson}
and on the coma expansion velocities of 17P/Holmes
\citep{2008A&A...479L..45Montalto, 2010MNRAS.407.1784Hsieh} and
C/Hale-Bopp \citep{2002EM&P...90....5Biver} vary from a few 100 m
s$^{-1}$ to 1000 m s$^{-1}$. We take an intermediate $v_d=500$ m
s$^{-1}$ and obtain an annulus crossing time \begin{equation}
  t_\mathrm{cross}=\frac{w_\mathrm{cross}}{v_d}\sim0.8\times10^4\;\mathrm{s}.
\end{equation} Finally, the mass loss rate is given by
$dM_d/dt=M_d/t_\mathrm{cross}\sim1.1\times10^2$ kg/s. At this rate, a
nucleus of radius $r_e=3$ km (see \S \ref{Sec.NucleusSize}) and bulk
density $\rho_\mathrm{nuc}=1000$ kg m$^{-3}$ would have a sublimation
lifetime $t_\mathrm{sub}=3\times10^4$ yr, significantly shorter than
the dynamical lifetime of \LG\ (see \S \ref{Sec.OrbitHistory}).

\begin{figure}
  \includegraphics[width=83mm]{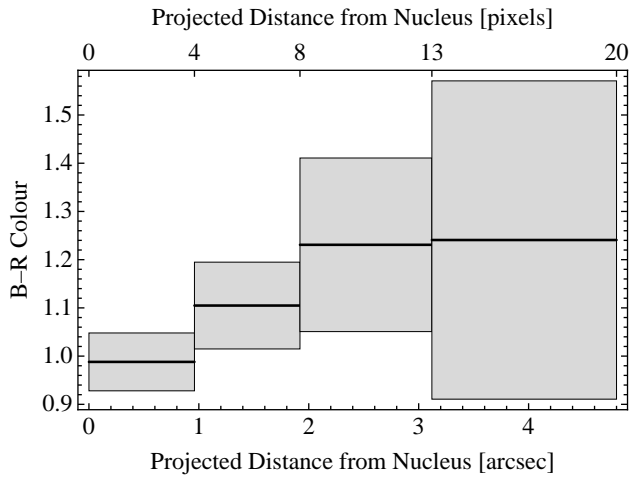}
  \centering

  \caption{Radial variation of the $B-R$ colour (thick horizontal
    lines) and uncertainty (grey boxes).  The average colours (thick
    horizontal lines) are shown in each of the four concentric regions
    described in Table \ref{Table.PhotometryRegions}. The
  coma-dominated annuli (apertures $>$ 8 pixels) appear redder than
the central region despite the large uncertainties.}

  \label{Fig.ColourVsDistance} 
\end{figure}

\begin{figure}
  \includegraphics[width=78mm]{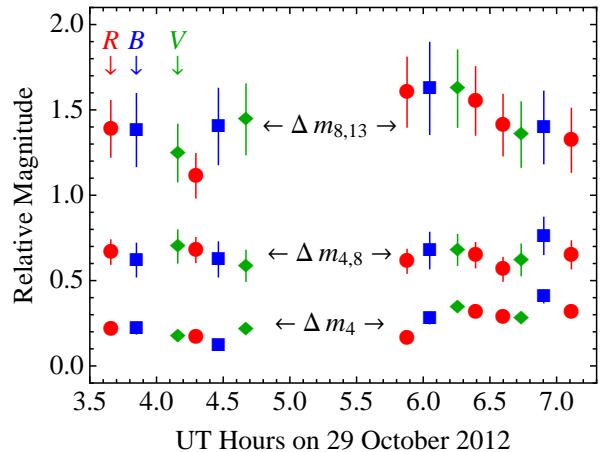}
  \centering

  \caption[FigLightcurve] {Lightcurves of comet \LG\ sampling the
    central region $\mathcal{R}_4$ ($\Delta m_4$, measured within a 4
    pixel or 0.96\arcsec\ radius aperture), the intermediate region
    $\mathcal{R}_{4,8}$ ($\Delta m_{4,8}$, from 0.96 to 1.92\arcsec),
    and the coma-dominated region $\mathcal{R}_{8,13}$ ($\Delta
    m_{8,13}$, from 1.92 to 3.12\arcsec). For each region, we plot
    relative magnitude lightcurves in bands $B$ (blue squares), $V$
    (green diamonds) and $R$ (red circles). The lightcurves in
    different bands were aligned simply by median subtraction, and the
    lightcurves in each region ($\Delta m_4$, $\Delta m_{4,8}$,
  $\Delta m_{8,13}$) were shifted vertically for clarity of
presentation.  }

  \label{Fig.Lightcurve} 
\end{figure}

\subsection{Nucleus and Coma Colour}\label{Sec.NucleusColour}

We investigated the possibility that the nucleus and coma dust of \LG\
have different colours by comparing the fluxes through different
filters within the regions detailed in Table
\ref{Table.PhotometryRegions}. To do so, we generated $B$- and
$V$-band stacks as described in \S\ref{Sec.NucleusSize} using the
frames with best image quality ($6\times B$-band, seeing 1.0\arcsec\
to 1.1\arcsec; $6\times V$-band, seeing 0.8\arcsec\ to 0.9\arcsec) to
obtain single images with equivalent exposure times 1440 s in $B$ (see
Fig.\ \ref{Fig.Stacks}) and 720 s in $V$.  The resulting stacked
images have $\mathrm{SNR}_B\sim32$ and $\mathrm{SNR}_V\sim34$ within
the central aperture.

Figure \ref{Fig.ColourVsDistance} plots the $B-R$ colour of \LG\
versus projected distance from the nucleus, and Table
\ref{Table.Photometry} lists the $B$, $V$ and $R$ apparent magnitudes
in the regions of interest.  The comet nucleus region,
$\mathcal{R}_4$, has colours $B-R=0.99\pm0.06$ mag and
$V-R=0.47\pm0.06$ mag, typical of Jupiter family comet nuclei
\citep{2002AJ....123.1039Jew, 2006MNRAS.373.1590Snodgrass}.  Further
from the nucleus, as the flux becomes dominated by the coma, the $B-R$
colour becomes slightly redder.  Despite the large uncertainties the
effect is noticeable and suggests that the coma dust may have a
different colour from the nucleus.

\subsection{Lightcurves}

We searched for photometric variability in the \LG\ data using the
individual frames taken in $B$, $V$ and $R$ over the course of the
night. Periodic photometric variations from the central region can
signal a rotating, elongated nucleus, while temporal variations in the
outer coma regions may indicate variable activity from the comet. To
look for photometric variations we performed photometry within the
regions listed in Table \ref{Table.PhotometryRegions} for each frame.
As part of our observing strategy we acquired sets of three
consecutive images per filter. That allowed to take the median
magnitude of each set and divide the median uncertainty by $\sqrt{3}$. 

Figure\ \ref{Fig.Lightcurve} shows the resulting $B$, $V$ and $R$
lightcurves in regions $\mathcal{R}_4$, $\mathcal{R}_{4,8}$ and
$\mathcal{R}_{8,13}$. Since we are only interested in the relative
variability, the lightcurves for each band were median subtracted for
alignment purposes. The lightcurves within each region were then
shifted vertically for clarity of presentation.

The small variability seen in $B$ and $R$ within the central region
$\mathcal{R}_4$ is probably not significant:  the magnitudes $m_4$ in
$B$ and $R$ display a positive correlation with atmospheric seeing, at
a level that has a probability of only 4\% of occurring by chance. The
large scale variation observed in $m_{8,13}$, in all three bands, does
not correlate with seeing nor airmass and is probably real. Although
the formal photometric uncertainties are large, the variation is
regular.  However, our time base is short rendering the interpretation
difficult.

\section{Dynamical Past and Future} \label{Sec.Dynamics}

Discovered by the LINEAR survey in October 2010, \LG\ was initially
classified as a Jupiter Trojan. The object retained its Trojan status
within the JPL Horizons system until October 2011, when cometary
activity was detected \citep{2011IAUC.9235....1Grauer}, at which point
it was reclassified as a JFC.  Currently, \LG\ has a low inclination
($i=2\fdg6$), nearly circular orbit ($e=0.09$) with a semimajor axis
$a=5.6$ au. The difference between the distances to perihelion and
aphelion is a mere 1 au, leading to a surface temperature variation of
$\Delta T=15$ K (blackbody in equilibrium with sunlight). Such a
temperature difference seems too small to cause periodic activity
close to perihelion as is the case for most JFCs.  This issue is
discussed in more detail in \S\ref{Sec.Discussion}.

\begin{figure}
  \includegraphics[width=83mm]{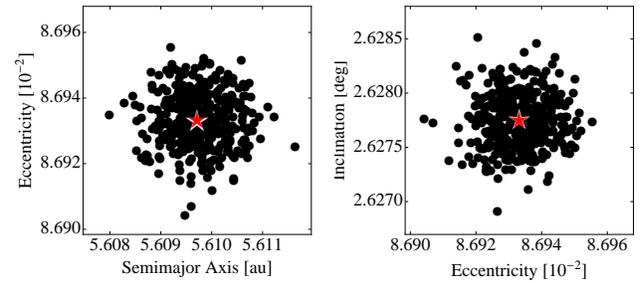}
  \centering

  \caption[FigCloneDistribution] {Initial orbital parametres of 370
    clones of \LG\ used in our numerical simulations. The clones are
    drawn from Gaussian distributions with centers on the actual
    orbital elements of \LG\ (star symbol) and standard deviations
  equal to one thousandth of the central value.}

  \label{Fig.CloneDistribution} 
\end{figure}

To investigate the past and future dynamical evolution of \LG, we
numerically integrate its current orbit backwards and forwards in
time.  For this purpose, we use the N-body integration package {\sc
Mercury} version 6.2 \citep{1999MNRAS.304..793Chambers}. We integrate
the orbits of 371 objects, \LG\ plus 370 clones with 
normally distributed orbital elements centered on \LG's current orbit
and with a 1-$\sigma$ dispersion equal to $10^{-4}$ of each parametre
(Fig.\ \ref{Fig.CloneDistribution}). The Sun and 8 major planets are
included as massive bodies and the 371 \LG\ clones as massless test
particles.  We use {\sc Mercury}'s hybrid algorithm mode which
combines a second order mixed-variable symplectic algorithm with a
Bulirsch-Stoer integrator to handle close encounters.  We select an
initial timestep of 8 days and remove clones that go beyond 200 au
from the Sun at any point during the dynamical evolution. 

We note that our simulations of the orbital evolution of \LG\ neglect
non-gravitational forces. Collimated sublimation jets accelerate the
nucleus and may alter the evolution presented here. An upper limit to
this effect can be calculated assuming that the non-gravitational
acceleration, $T$, is due to a single sublimation jet, always aligned
tangentially along the orbit of the comet, through which all mass loss
occurs. In that case, the rate of change of the semimajor axis,
$da/dt$, can be written as \begin{equation}
da/dt=2\,V\,a^2\,T/\left(G\,M_{\sun}\right) \end{equation} where $V$
is the orbital velocity, $G$ is the gravitational constant, $M_{\sun}$
is the mass of the Sun, and the acceleration due to the jet is
\begin{equation}
T=\left(dM_d/dt\right)\left(v_d/m_\mathrm{nuc}\right).  \end{equation}
Substituting $dM_d/dt=10^2$ kg, $v_d=500$ m s$^{-1}$,
$m_\mathrm{nuc}=1.1\times10^{14}$ kg, and taking $V$ as being the
circular velocity at $a=5.6$ au we obtain $da/dt=1.4\times10^{-5}$ AU
yr$^{-1}$. In the sublimation lifetime ($t_\mathrm{sub}=3\times10^4$
yr) of comet \LG\ its orbit can decay by $\left(da/dt\right)\times
t_\mathrm{sub}=0.42$ au. This value is a strong upper limit, as it
assumes a jet geometry contrived to achieve maximum orbital decay.

\subsection{Orbit Stability}
\label{Sec.OrbitStability}

Our simulations show that \LG\ is dynamically unstable on very short
timescales.  The 1-$\sigma$ spread of the main orbital parametres
(semimajor axis, eccentricity, inclination) of the cloud of \LG\ clone
particles expands by a factor $e$ in only 12 to 14 yr into the past
and about 100 yr into the future. These times provide a reasonable
estimate of the Lyapunov timescale for \LG. 

Due to the strongly chaotic nature of the orbit of \LG, we can expect
at best to obtain a statistical assessment of the past and future
evolution of this comet. The chaotic evolution also implies that long
term integrations into the past or future are statistically equivalent
\citep{1994Icar..108...18L, 2004MNRAS.354..798Horner}. However,
current close encounters between \LG\ and Jupiter (see
\S\ref{Sec.CloseEncounter}) generate an asymmetry in the dynamical
evolution making it interesting to investigate the short term past and
future cases separately.  Those same close encounters lead to further
strong divergence of the clone orbits reinforcing the need for a more
statistical analysis beyond a few kyr from the current time.  

\begin{figure*} 
  \includegraphics[width=150mm]{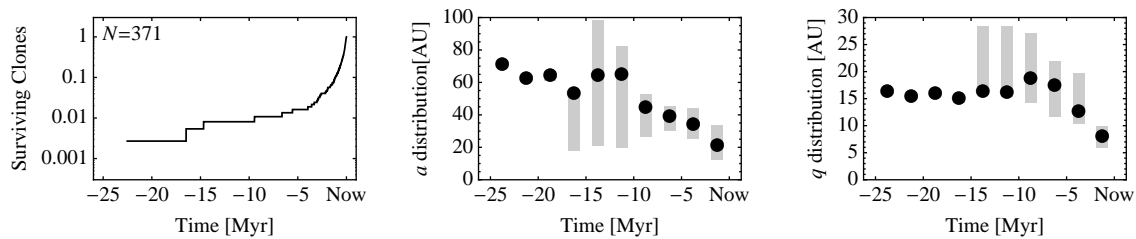}
  \centering

  \caption[FigPastOrbitalEvolution] {Past orbital evolution of the 371
    clones of \LG. The fraction of surviving clones (left) decreases
    steeply as their orbits are integrated back in time.  Half the
    clones do not survive the orbital integration beyond $t=-182$ kyr.
    The longest surviving clone reaches $-22.5$ Myrs.  The surviving
    clones evolve into Centaur/Scattered Kuiper Belt Object orbits
    with increasing median semimajor axis (black points in center
    plot) and median perihelion distance (black points in plot on the
    right). The gray boxes contain the central 50\% of the distribution in
  each time bin.}

  \label{Fig.PastOrbitalEvolution} 
\end{figure*} 

\subsection{Orbit History} \label{Sec.OrbitHistory}

When integrating into the past we find that 50\% of the clones are
removed from the simulation by $t=-182$ kyr. The longest lived clone
reaches $t=-22.5$ Myr. The fraction of surviving clones is plotted
versus time in Fig.\ \ref{Fig.PastOrbitalEvolution} (left panel). Each
step in the figure corresponds to the removal of 1 clone, so, e.g.\
only 3 clones survive to times earlier than $t=-10$ Myr. The Figure
also shows the median semimajor axis and perihelion distance of the
surviving clones. The former increases with time into the past, albeit
with large scatter (indicated by the gray boxes which contain the
central 50\% of the points). The latter increases to about $q\sim17$
au and then remains approximately constant with a tail towards larger
$q$.  The past evolution suggests that \LG\ has recently evolved into
its current orbit from a Centaur-type orbit.

\begin{figure} 
  \includegraphics[width=83mm]{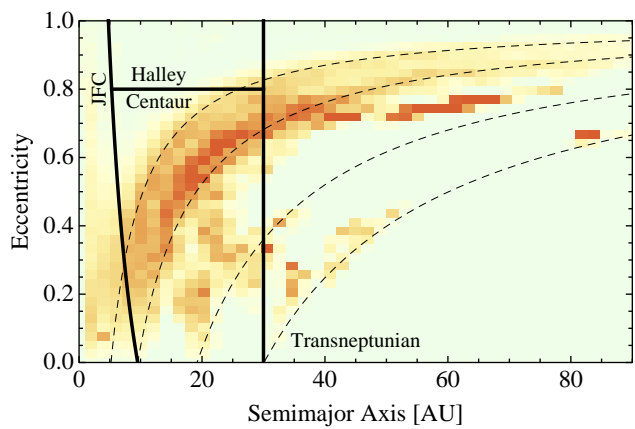}
  \centering

  \caption[OrbitHistory] {Heat map of the evolution of \LG\ clones
    during the backwards integration described in
    \S\ref{Sec.Dynamics}.  Darker patches have been occupied for a
    longer time by clones. The $a-e$ space is divided into 4 regions,
    representative of Jupiter family comets: $a<a_S/(1+e)$, Centaurs:
    $a_S/(1+e)<a<a_N$ and $e<0.8$, Halley comets: Centaurs with
    $e>0.8$, and Transneptunian objects: $a>a_N$, where $a_S$ and
    $a_N$ are the semimajor axes of Saturn and Neptune.  Dashed lines
    mark the perihelia of Jupiter, Saturn, Uranus and Neptune. The
    clones spend most of the time as Centaurs, although pathways
  exist that transport clones to the transneptunian region.}

  \label{Fig.OrbitHistory} 
\end{figure} 

\begin{figure} 
  \includegraphics[width=83mm]{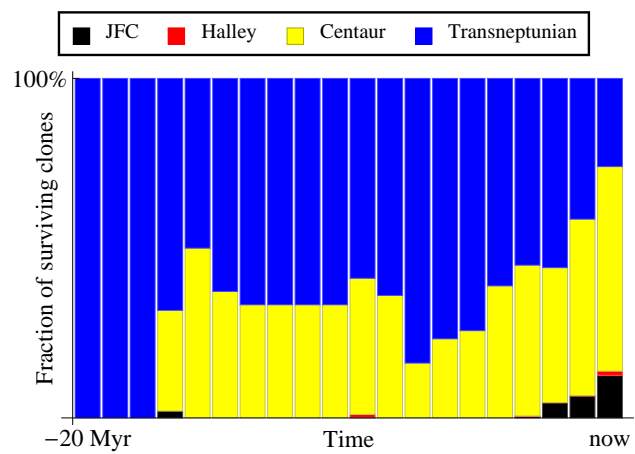}
  \centering

  \caption[FigOrbitTypeHistory] {Fraction of surviving clones within
    each of the regions (orbit types) described in Fig.\
    \ref{Fig.OrbitHistory} plotted against time. Bins are 1 Myr wide.
    The leftmost three bins ($-17$ Myr to $-20$ Myr) contain a single
  clone (see Fig.\ \ref{Fig.PastOrbitalEvolution}).}

  \label{Fig.OrbitTypeHistory} 
\end{figure}

Figure\ \ref{Fig.OrbitHistory} shows a heat map of the evolution of
the \LG\ clones in the semimajor axis vs.\ eccentricity plane.
Darker/redder regions are more often visited by clones as their orbits
evolve backwards in time. The plane is divided into four regions,
taken roughly to represent four types of orbits: {\em JFC-type},
defined as having $a<a_S/(1+e)$, {\em Centaur-type}, bound by
$a_S/(1+e)<a<a_N$ and $e<0.8$, {\em Halley-type}, similar to
Centaur-type except with $e>0.8$, and {\em transneptunian-type}, with
$a>a_N$, where $a_S$ and $a_N$ are the semimajor axes of Saturn and
Neptune.  These regions do not aim at accurately representing the
dynamical behaviour of clone particles; they are designed simply as a
means to classify the type of orbit occupied by the clones as a
function of time. The Figure shows that clones evolve mainly along
constant-perihelion lines, as their semimajor axis and eccentricity
increase. Figure\ \ref{Fig.OrbitTypeHistory} traces the history of
clones across the $a$-$e$ plane as a function of time. The Figure
shows the fraction of surviving clones within each of the orbit-type
regions described above in bins of 1 Myr, 20 Myrs into the past. As
already seen in Fig.\ \ref{Fig.PastOrbitalEvolution}, the surviving
clones tend to quickly move away from the JFC region into Centaur-type
orbits. Beyond just a few Myrs into the past the semimajor axis of the
surviving clones moves even further away from the sun, into the
transneptunian region. The last surviving clone possesses a semimajor
axis and eccentricity typical of a scattered transneptunian object
($a\sim80$ au, $e\sim0.65$).

\begin{figure} 
  \centering
  \includegraphics[width=83mm]{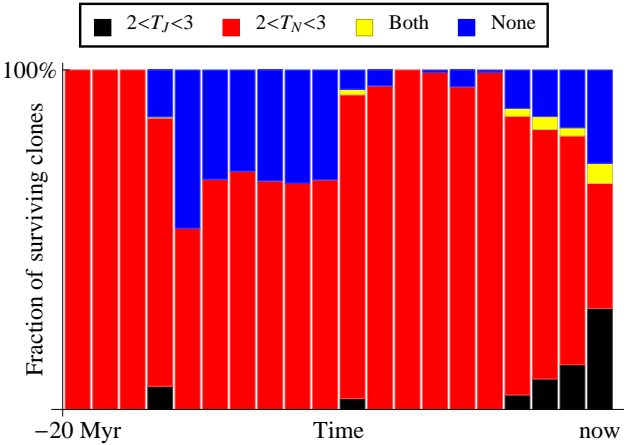}
  \centering

  \caption[FigTisserandHistory] {Evolution of the Tisserand parametre
    of clones. Myr-wide bins show the fractions of clones that have
    Tisserand parametre $2<T\leq3$ with respect to Jupiter, Neptune,
    both planets, or none. A value $2<T\leq3$ with respect to a given
    planet indicates that that planet dominates the orbit of the object.
    The Figure shows that, as time evolves into the past, Jupiter hands
  control to Neptune as main influence to the clone orbits.}

  \label{Fig.TisserandHistory} 
\end{figure}

In Figure\ \ref{Fig.TisserandHistory} we show the history of the
Tisserand parametre of \LG\ clones with respect to Jupiter and Neptune
in our numerical simulation. The Tisserand parameter is a useful
dynamical quantity that is conserved in the restricted three body
problem. Here, the three bodies are the Sun, a planet and \LG. In the
solar system, the Tisserand parameter is not exactly conserved, but
when calculated for a given pair of small body and planet it can be
used to quantify the dynamical influence the latter has on the former.
For instance, the JFCs have Tisserand parametres $2<T_J<3$ with
respect to Jupiter, to which their orbits are strongly dynamically
coupled.  Main belt asteroids have $T_J>3$ and are relatively stable
with respect to the Jovian gas giant.  The Tisserand parametre of \LG,
calculated with respect to planet $P$, is given by \begin{equation}
T_P=a_P/a+2\sqrt{(1-e^2)(a/a_P)}\cos i \end{equation} \noindent where
$a_P$ is the semimajor axis of the planet and $a$, $e$ and $i$ are the
semimajor axis, eccentricity and inclination of \LG.  As shown in
Figure\ \ref{Fig.TisserandHistory}, within the first 1 Myr into the
past the clones of \LG\ are controlled partly by Jupiter ($2<T_J<3$),
partly by Neptune ($2<T_N<3$) and partly by neither of those two
planets.  The latter indicates orbits between Jupiter and Neptune.
Further into the past, Jupiter becomes less important dynamically
while Neptune becomes more dominant.

In summary, our simulation results conspire to suggest that \LG\ may
be a recent arrival in the inner solar system, possibly from a Centaur
or transneptunian type orbit. That makes this object an interesting
target of study as it may offer a glimpse of a relatively unprocessed
Centaur.

\subsection{Close Encounters with Jupiter and the\\* Trojan Clouds}
\label{Sec.CloseEncounter}

We performed a high time-resolution simulation (0.1 yr initial
timestep) going back only 3000 yr to investigate recent close
encounters between the \LG\ clones and the major planets. We focus
here on Jupiter, due to its current close proximity to \LG.  In the
very near past, we find that of the 371 \LG\ clones, 331 (89\%) had
close encounters with Jupiter ($<100$ planet radii) between 48 and 51
yr ago. During that time interval, 34 clone particles came within 20
Jupiter radii ($r_J$) of the giant planet and the closest approach
distance for any particle was $13\,r_J$. Two of the 331 clone
particles had very close approaches to Jupiter in consecutive orbits:
one particle came within $14.0\,r_J$ just 48 yr ago and within
$8.7\,r_J$ during the previous orbit (60 yr ago); another particle
passed at $16.0\,r_J$ and $14.6\,r_J$ about 48 and 57 yr ago.
Finally, if we consider the full 3000 yr time span of the simulation,
7 particles came within the Roche limit of Jupiter
\citep[$\approx2.7r_J$ assuming a fluid-like \LG\ nucleus with bulk
density $\rho_\mathrm{nuc}=1000$ kg
m$^{-3}$;][]{1996Icar..121..225Asphaug}. 
In a similar integration to the one just described but into the future
we find that 11 out of the 371 clone particles come within 10 $r_J$
and 3 of those come within 2 $r_J$ roughly 145 yr into the future.  A
total of 41 clone particles approach Jupiter within 100 $r_J$ at that
time.

We also searched for possible crossings of one of the Jovian Trojan
clouds by \LG\ clones. The Trojan clouds have an irregular shape,
roughly $4\times2\times0.5$ au (FWHM) in size
\citep{2000AJ....120.1140Jewitt, 2008PASJ...60..293Nakamura}. To
identify potential crossings and estimate the time spent by clones
inside the Trojan clouds we used Jupiter's position to calculate the
positions of its L3 and L4 Lagrangian points in each timestep.
Finally, for each clone we added the timesteps it spent within 0.3 au
of either L3 or L4. We chose 0.3 au to ensure the clone is well within
the densest region of the Trojan clouds. We found that clones spend an
average of 10.1 of the last 3000 yr within the regions just described.

\begin{table*}
  \centering
  \begin{minipage}{140mm}
    \caption{Comparison between \LG\ and 29P \label{Table.Comparison}}
    \begin{tabular}{@{}lccccccccc@{}}
      \hline
      Comet   & $a$ & $e$ & $i$ & $B-R$ & $V-R$ & $r_e$ & $M$ & $dM/dt$ & $dm/dt$ \\
      & [au]    &  & [\degr] & [mag] & [mag] & [km] & [kg] & [kg s$^{-1}$] & [kg m$^{-2}$ s$^{-1}$] \\ 
    \hline
    \LG  & 5.6 & 0.09 & 2.6 & $0.99\pm0.06$ & $0.47\pm0.06$ & $<3$     & $1.1\times10^{14}$ &
    $1.1\times10^2$ & $7.7\times10^{-7}$ \\
    29P  & 6.0 & 0.04 & 9.4 & $1.28\pm0.04$ & $0.50\pm0.03$ & $23\pm3$ & $5.1\times10^{16}$ &
    $5.1\times10^3$ & $9.7\times10^{-7}$ \\
    \hline
  \end{tabular}

  \medskip Columns are (1) comet ID, (2) semimajor axis, (3) eccentricity and
  (4) inclination, (5) and (6) broadband colours, (7) equivalent radius, (8)
  model dependent mass-loss rate, (9) specific mass-loss rate.  Mass assumes
  nucleus/dust bulk density $\rho_\mathrm{nuc}=1000$ kg m$^{-3}$.  Specific
  mass-loss rate assumes a spherical nucleus of radius $r_e$. Comet 29P colours
  from \citet{2009AJ....137.4296Jewitt}.
\end{minipage}
\end{table*}

\section{Discussion} \label{Sec.Discussion}

Figure \ref{Fig.PSOne} shows \LG\ as seen by the PanSTARRS PS1 survey
(see \S\ref{Sec.Observations}) close to the time the comet was
discovered by LINEAR. The PS1 images show that \LG\ was active at the
time of discovery and yet the LINEAR pipeline first classified this
object as a Jovian Trojan. This suggests that LINEAR may have
misclassified several active objects as inert asteroids, which is an
important point to consider at a time when the study of active
asteroids is receiving increasing interest \citep[][and references
therein]{2012AJ....143...66Jewitt}.

Our data reveal that \LG\ was active in October 2010, roughly 60\degr\
past perihelion, and remains active a year later, at 90\degr\ past
perihelion. \LG's nearly circular orbit, atypical for a JFC, leads us
to consider a number of explanations for its activity. One possibility
is that \LG\ has been activated by a recent collision with a smaller
object. We find this unlikely: our simulations show that \LG\ spent
only a negligible fraction of its recent past within the Trojan clouds
(but see below), and even there the chance of a collision onto its
$r_e<3$ km nucleus would be low \citep[intrinsic collision probability
$P_i=6.5\times 10^{-18}$ km$^{-2}$
yr$^{-1}$;][]{1998A&A...339..272Delloro}. Another possibility is that
\LG\ was exposed to significant tidal stress due to close approaches
to Jupiter. This scenario would require encounters with Jupiter closer
than the Roche limit \citep[mass shedding begins at
$d<0.69R_\mathrm{roche}$ for a comet on a parabolic
orbit;][]{1992Icar...95...86Sridhar,1996Icar..121..225Asphaug} which
our simulations show to be improbable. It is also possible that \LG\
became active due to rotational mass shedding but since the rotation
period of \LG\ is not known this possibility is untestable.  One
scenario that our simulations can not rule out is that \LG\ originated
in the Trojan population and was disloged from the 1:1 resonance with
Jupiter through the action of non-gravitational effects (see
\S\ref{Sec.Dynamics}). Conceivably, \LG\ could have been collisionally
actived by a smaller Trojan and led to drift from the stable region,
accelerated by mass-loss jets; we note that \LG's highly unstable
orbit implies a recent departure from the Trojan region. Modelling
this possibility is a complex task involving a number of unknown
parametres and is beyond the scope of this paper but future
observations of this comet may shed light on a possible link with the
Trojan population.

The activity of \LG\ could simply be due to ice sublimation. At first
sight, the small difference between the \LG\ perihelion and aphelion
distances (Table \ref{Table.Elements}), and the consequently small
orbital variation of the subsolar equilibrium temperature ($\Delta
T\approx15$ K) could seem insufficient to power periodic sublimation
activity. However, between the perihelion and aphelion distances of
\LG, the specific mass loss rate due to water ice sublimation varies
by more than an order of magnitude (from $10^{-5}$ to $10^{-4}$ kg
m$^{-2}$ s$^{-1}$) while CO/CO$_2$ sublimation driven mass loss varies
by a factor of 2 and is around $10^{-2}$ kg m$^{-2}$ s$^{-1}$
\citep[][]{2009AJ....137.4296Jewitt}.  We find that \LG\ loses mass
into the coma at a rate $\sim10^{2}$ kg s$^{-1}$ implying that an area
of $10^6$ to $10^7$ m$^2$ would need to be active on the surface of
\LG\ if water ice sublimation is the source of the activity. Those
areas are small compared to the maximum surface area of the nucleus,
$4\pi r_e^2\sim10^8$ m$^2$.  Sublimation due to CO/CO$_2$ would
require only an area $\sim10^4$ m$^2$ to be active, corresponding to a
tiny active vent on the surface of the 3 km nucleus. We conclude that
given the recent perihelion passage and the particular range of
heliocentric distances traversed by \LG\ since then our estimated
current mass loss rate is consistent with the activity being driven by
ice sublimation. 

\begin{figure} 
  \includegraphics[width=83mm]{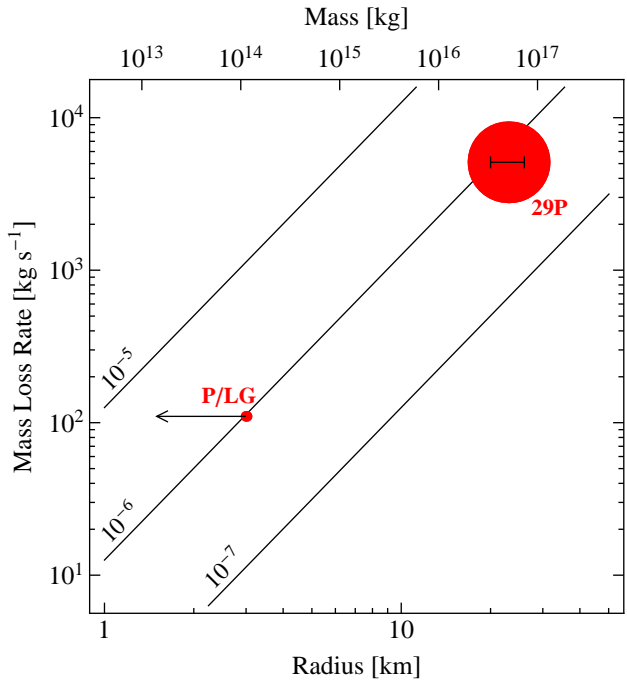}
  \centering

  \caption[FigLGSWComparison] {Comparison between comets \LG\ and 29P.
  Mass-loss rates are plotted against nucleus radius (and mass,
  assuming nucleus/dust bulk density $\rho_\mathrm{nuc}=1000$ kg
  m$^{-3}$). The arrow indicates that we possess only an upper limit
  on the nucleus radius of \LG. The uncertainty in the radius of 29P
  is marked by a horizontal error bar. The size of the \LG\ and 29P
  points is linearly proportional to their geometric cross-section for
  visual comparison. Lines of constant specific mass-loss rate are
  labeled in units of kg m$^{-2}$ s$^{-1}$.}

  \label{Fig.Comparison} 
\end{figure}

\LG\ is in many ways comparable to comet 29P (see Table
\ref{Table.Comparison} and Fig.\ \ref{Fig.Comparison}). Both comets
have high perihelion ($q>5$ au), low inclination ($i<10\degr$), nearly
circular orbits ($e<0.1$) more typical of Centaurs than JFCs. Both
orbits are unstable on very short timescales \citep[a few hundred
years;][]{2004MNRAS.354..798Horner}.  The two comets show similar
levels of activity, with specific mass-loss rates close to
$dm/dt=10^{-6}$ kg m$^{-2}$ s$^{-1}$, despite 29P being nearly an
order of magnitude larger ($>$2 orders of magnitude more massive) than
\LG.  Both objects display continued activity, although the coverage
of \LG\ reported here is not enough to establish a pattern of
activity. It will be interesting to see if \LG's activity is
significantly modulated by its orbital motion or if it remains
constant as is the case for comet 29P. It will also of interest to
monitor \LG\ and see if it too will display the sporadic outbursts
seen in 29P. Significant differences include the size of the comets
and the source of activity: 29P is dominated by CO sublimation, while
\LG\ is more likely active due to water-ice sublimation.

\section{Conclusions}

We report photometric observations and numerical simulations of the
orbital evolution of the unusual comet P/2010 TO20 LINEAR-Grauer
(\LG). Our findings can be summarised as follows:

\begin{enumerate}

\item Comet \LG\ was active at the time of discovery (October 2010) by
LINEAR, and remains active in October 2011. LINEAR did not detect the
activity and initially misclassified \LG\ as a Trojan, which suggests
that several active objects may have gone undetected by the survey.

\item The nucleus of \LG\ has equivalent radius $r_e<3$ km, and
colours $B-R=0.99\pm0.06$ mag and $V-R=0.47\pm0.06$ mag, values
typical of Jupiter family comets. The data suggest a slight reddening
of the dust colour with distance from the nucleus but the
uncertainties are large.  We find no significant rotational
photometric variability from the nucleus region.

\item We obtain a model-dependent estimate of the mass-loss rate from
\LG\ of $\sim$100 kg s$^{-1}$. We favour water-ice sublimation as the
simplest and most likely cause for activity in comet \LG. 

\item Our numerical simulations show that the orbit of \LG\ is
unstable on very short timescales and suggest that it may be a Centaur
that recently arrived in the inner solar system, although other
possibilities exist involving non-gravitational effects.

\item Comet \LG\ is in a number of ways reminiscent of the well-known
29P/Schwassmann-Wachmann 1. 29P is a comet/active Centaur that shows
sporadic outbursts superimposed on a background of constant activity.
Comet \LG\ is an order of magnitude smaller (three orders of magnitude
less massive) than 29P and yet displays similar activity per unit
area. Comets 29P and \LG\ are interesting as possible examples of
relatively unprocessed Centaurs.

\end{enumerate}

\section*{Acknowledgments}

I thank the referee, Mario Melita, as well as David Jewitt and Ivo
Labb\'e for their helpful comments on the manuscript. I am grateful to
Larry Denneau and Ken Smith for helpful assistance with the PS1 survey
data.  I also acknowledge Colin Snodgrass for kindly supplying an IRAF
fringe removal routine optimised for EFOSC2.

The PS1 Surveys have been made possible through contributions of the
Institute for Astronomy, the University of Hawaii, the Pan-STARRS
Project Office, the Max-Planck Society and its participating
institutes, the Max Planck Institute for Astronomy, Heidelberg and the
Max Planck Institute for Extraterrestrial Physics, Garching, The Johns
Hopkins University, Durham University, the University of Edinburgh,
Queen's University Belfast, the Harvard-Smithsonian Center for
Astrophysics, and the Las Cumbres Observatory Global Telescope
Network, Incorporated, the National Central University of Taiwan, and
the National Aeronautics and Space Administration under Grant No.
NNX08AR22G issued through the Planetary Science Division of the NASA
Science Mission Directorate.

\bsp

\label{lastpage}


\begin{thebibliography}{99}
\bibitem[{{Asphaug} \& {Benz}(1996)}]{1996Icar..121..225Asphaug}
{Asphaug} E., {Benz} W., 1996, Icarus, 121, 225

\bibitem[{{Biver} {et~al}\mbox{.}(2002){Biver}, {Bockel{\'e}e-Morvan}, {Colom},
  {Crovisier}, {Henry}, {Lellouch}, {Winnberg}, {Johansson}, {Gunnarsson},
  {Rickman}, {Rantakyr{\"o}}, {Davies}, {Dent}, {Paubert}, {Moreno}, {Wink},
  {Despois}, {Benford}, {Gardner}, {Lis}, {Mehringer}, {Phillips}, \&
  {Rauer}}]{2002EM&P...90....5Biver}
{Biver} N. {et~al.}, 2002, Earth Moon and Planets, 90, 5

\bibitem[{{Buzzoni} {et~al}\mbox{.}(1984){Buzzoni}, {Delabre}, {Dekker},
  {Dodorico}, {Enard}, {Focardi}, {Gustafsson}, {Nees}, {Paureau}, \&
  {Reiss}}]{1984Msngr..38....9Buzzoni}
{Buzzoni} B. {et~al.}, 1984, The Messenger, 38, 9

\bibitem[{{Chambers}(1999)}]{1999MNRAS.304..793Chambers}
{Chambers} J.~E., 1999, MNRAS, 304, 793

\bibitem[{{Crifo} {et~al}\mbox{.}(2004){Crifo}, {Fulle}, {K{\"o}mle}, \&
  {Szego}}]{2004come.book..471Crifo}
{Crifo} J.~F., {Fulle} M., {K{\"o}mle} N.~I., {Szego} K., 2004, {Nucleus-coma
  structural relationships: lessons from physical models}, {Festou} M.~C.,
  {Keller} H.~U., {Weaver} H.~A., eds., pp. 471--503

\bibitem[{{Crovisier} {et~al}\mbox{.}(1995){Crovisier}, {Biver},
  {Bockelee-Morvan}, {Colom}, {Jorda}, {Lellouch}, {Paubert}, \&
  {Despois}}]{1995Icar..115..213Crovisier}
{Crovisier} J., {Biver} N., {Bockelee-Morvan} D., {Colom} P., {Jorda} L.,
  {Lellouch} E., {Paubert} G., {Despois} D., 1995, Icarus, 115, 213

\bibitem[{{dell'Oro} {et~al}\mbox{.}(1998){dell'Oro}, {PAolicchi}, {Marzari},
  {Dotto}, \& {Vanzani}}]{1998A&A...339..272Delloro}
{dell'Oro} A., {PAolicchi} F., {Marzari} P., {Dotto} E., {Vanzani} V., 1998,
  A\&A, 339, 272

\bibitem[\protect\citeauthoryear{Grauer et
al.}{2011}]{2011IAUC.9235....1Grauer} 
Grauer A.~D., Sostero G., Melville I., Kasprzyk A., Howes N., Guido E., 
Spahr T., Williams G.~V., 2011, IAUC, 9235, 1 

\bibitem[{{Horner}, {Evans} \& {Bailey}(2004){Horner}, {Evans}, \&
  {Bailey}}]{2004MNRAS.354..798Horner}
{Horner} J., {Evans} N.~W., {Bailey} M.~E., 2004, MNRAS, 354, 798

\bibitem[{{Hsieh} {et~al}\mbox{.}(2010){Hsieh}, {Fitzsimmons}, {Joshi},
  {Christian}, \& {Pollacco}}]{2010MNRAS.407.1784Hsieh}
{Hsieh} H.~H., {Fitzsimmons} A., {Joshi} Y., {Christian} D., {Pollacco} D.~L.,
  2010, MNRAS, 407, 1784

\bibitem[{{Jewitt}(1990)}]{1990ApJ...351..277Jewitt}
{Jewitt} D., 1990, ApJ, 351, 277

\bibitem[{{Jewitt}(2009)}]{2009AJ....137.4296Jewitt}
{Jewitt} D., 2009, AJ, 137, 4296

\bibitem[{{Jewitt}(2012)}]{2012AJ....143...66Jewitt}
{Jewitt} D., 2012, AJ, 143, 66

\bibitem[{{Jewitt} \& {Luu}(1989)}]{1989AJ.....97.1766Jewitt}
{Jewitt} D., {Luu} J., 1989, AJ, 97, 1766

\bibitem[{{Jewitt}(2002)}]{2002AJ....123.1039Jew}
{Jewitt} D.~C., 2002, AJ, 123, 1039

\bibitem[{{Jewitt} \& {Luu}(2001)}]{2001AJ....122.2099J}
{Jewitt} D.~C., {Luu} J.~X., 2001, AJ, 122, 2099

\bibitem[{{Jewitt}, {Trujillo} \& {Luu}(2000){Jewitt}, {Trujillo}, \&
  {Luu}}]{2000AJ....120.1140Jewitt}
{Jewitt} D.~C., {Trujillo} C.~A., {Luu} J.~X., 2000, AJ, 120, 1140

\bibitem[{{Kolokolova} {et~al}\mbox{.}(2004){Kolokolova}, {Hanner},
  {Levasseur-Regourd}, \& {Gustafson}}]{2004come.book..577Kolokolova}
{Kolokolova} L., {Hanner} M.~S., {Levasseur-Regourd} A.-C., {Gustafson}
  B.~{\AA}.~S., 2004, {Physical properties of cometary dust from light
  scattering and thermal emission}, {Festou} M.~C., {Keller} H.~U., {Weaver}
  H.~A., eds., pp. 577--604

\bibitem[{{Lamy} \& {Toth}(2009)}]{2009Icar..201..674Lamy}
{Lamy} P., {Toth} I., 2009, Icarus, 201, 674

\bibitem[{{Landolt}(1992)}]{1992AJ....104..340Lan}
{Landolt} A.~U., 1992, AJ, 104, 340

\bibitem[{{Levison} \& {Duncan}(1994)}]{1994Icar..108...18L}
{Levison} H.~F., {Duncan} M.~J., 1994, Icarus, 108, 18

\bibitem[{{Li} {et~al}\mbox{.}(2011){Li}, {Jewitt}, {Clover}, \&
  {Jackson}}]{2011ApJ...728...31Li}
{Li} J., {Jewitt} D., {Clover} J.~M., {Jackson} B.~V., 2011, ApJ, 728, 31

\bibitem[{{Meech} \& {Jewitt}(1987)}]{1987A&A...187..585Meech}
{Meech} K.~J., {Jewitt} D.~C., 1987, A\&A, 187, 585

\bibitem[{{Melita} \& {Licandro}(2012)}]{2012A&A...539A.144Melita}
{Melita} M.~D., {Licandro} J., 2012, A\&A, 539, A144

\bibitem[{{Millis}, {Ahearn} \& {Thompson}(1982){Millis}, {Ahearn}, \&
  {Thompson}}]{1982AJ.....87.1310Millis}
{Millis} R.~L., {Ahearn} M.~F., {Thompson} D.~T., 1982, AJ, 87, 1310

\bibitem[{{Montalto} {et~al}\mbox{.}(2008){Montalto}, {Riffeser}, {Hopp},
  {Wilke}, \& {Carraro}}]{2008A&A...479L..45Montalto}
{Montalto} M., {Riffeser} A., {Hopp} U., {Wilke} S., {Carraro} G.,
2008, A\&A,
  479, L45

\bibitem[{{Moroz} {et~al}\mbox{.}(1998){Moroz}, {Arnold}, {Korochantsev}, \&
  {Wasch}}]{1998Icar..134..253Moroz}
{Moroz} L.~V., {Arnold} G., {Korochantsev} A.~V., {Wasch} R., 1998, Icarus,
  134, 253

\bibitem[\protect\citeauthoryear{Nakamura \&
Yoshida}{2008}]{2008PASJ...60..293Nakamura} Nakamura T., Yoshida F., 2008,
PASJ, 60, 293 

\bibitem[{{Peixinho} {et~al}\mbox{.}(2012){Peixinho}, {Delsanti},
  {Guilbert-Lepoutre}, {Gafeira}, \& {Lacerda}}]{2012arXiv1206.3153Peixinho}
{Peixinho} N., {Delsanti} A., {Guilbert-Lepoutre} A., {Gafeira} R., {Lacerda}
  P., 2012, A\&A (in press)

\bibitem[{{Peixinho} {et~al}\mbox{.}(2003){Peixinho}, {Doressoundiram},
  {Delsanti}, {Boehnhardt}, {Barucci}, \& {Belskaya}}]{2003A&A...410L..29Pei}
{Peixinho} N., {Doressoundiram} A., {Delsanti} A., {Boehnhardt} H., {Barucci}
  M.~A., {Belskaya} I., 2003, A\&A, 410, L29

\bibitem[\protect\citeauthoryear{Roemer}{1958}]{1958PASP...70..272Roemer}
Roemer E., 1958, PASP, 70, 272 

\bibitem[{{Russell}(1916)}]{1916ApJ....43..173Russell}
{Russell} H.~N., 1916, ApJ, 43, 173

\bibitem[\protect\citeauthoryear{Senay \&
Jewitt}{1994}]{1994Natur.371..229Senay} Senay M.~C., Jewitt D., 1994,
Natur, 371, 229 

\bibitem[{{Snodgrass}, {Lowry} \& {Fitzsimmons}(2006){Snodgrass}, {Lowry}, \&
  {Fitzsimmons}}]{2006MNRAS.373.1590Snodgrass}
{Snodgrass} C., {Lowry} S.~C., {Fitzsimmons} A., 2006, MNRAS, 373, 1590

\bibitem[{{Snodgrass} {et~al}\mbox{.}(2008){Snodgrass}, {Saviane}, {Monaco}, \&
  {Sinclaire}}]{2008Msngr.132...18Snodgrass}
{Snodgrass} C., {Saviane} I., {Monaco} L., {Sinclaire} P., 2008, The Messenger,
  132, 18

\bibitem[{{Solontoi} {et~al}\mbox{.}(2012){Solontoi}, {Ivezi{\'c}},
  {Juri{\'c}}, {Becker}, {Jones}, {West}, {Kent}, {Lupton}, {Claire}, {Knapp},
  {Quinn}, {Gunn}, \& {Schneider}}]{2012Icar..218..571Solontoi}
{Solontoi} M. {et~al.}, 2012, Icarus, 218, 571

\bibitem[{{Sridhar} \& {Tremaine}(1992)}]{1992Icar...95...86Sridhar}
{Sridhar} S., {Tremaine} S., 1992, Icarus, 95, 86

\bibitem[{{Stevenson}, {Kleyna} \& {Jewitt}(2010){Stevenson}, {Kleyna}, \&
  {Jewitt}}]{2010AJ....139.2230Stevenson}
{Stevenson} R., {Kleyna} J., {Jewitt} D., 2010, AJ, 139, 2230

\bibitem[\protect\citeauthoryear{Thompson et
al.}{1987}]{1987JGR....9214933Thompson} Thompson W.~R., Murray
B.~G.~J.~P.~T., Khare B.~N., Sagan C., 1987, JGR, 92, 14933 

\bibitem[{{Tonry} {et~al}\mbox{.}(2012){Tonry}, {Stubbs}, {Lykke}, {Doherty},
  {Shivvers}, {Burgett}, {Chambers}, {Hodapp}, {Kaiser}, {Kudritzki},
  {Magnier}, {Morgan}, {Price}, \& {Wainscoat}}]{2012ApJ...750...99Tonry}
{Tonry} J.~L. {et~al.}, 2012, ApJ, 750, 99

\bibitem[{{Trigo-Rodr{\'{\i}}guez}
  {et~al}\mbox{.}(2008){Trigo-Rodr{\'{\i}}guez}, {Garc{\'{\i}}a-Melendo},
  {Davidsson}, {S{\'a}nchez}, {Rodr{\'{\i}}guez}, {Lacruz}, {de Los Reyes}, \&
  {Pastor}}]{2008A&A...485..599TrigoRodriguez}
{Trigo-Rodr{\'{\i}}guez} J.~M., {Garc{\'{\i}}a-Melendo} E., {Davidsson}
  B.~J.~R., {S{\'a}nchez} A., {Rodr{\'{\i}}guez} D., {Lacruz} J., {de Los
  Reyes} J.~A., {Pastor} S., 2008, A\&A, 485, 599

\end{thebibliography}
\end{document}